\documentclass[dvips]{acta}
\usepackage{supertabular,lscape,epsfig}
\usepackage{amssymb}
\usepackage{amsmath}
\usepackage{multirow}
\usepackage{float}
\mathchardef\mhyphen="2D
\usepackage{placeins}
\DeclareSymbolFont{ppa}{OT1}{ppl}{m}{it}
\DeclareMathSymbol{\vv}{\mathalpha}{ppa}{'166}

\SetPages{1}{20}

\SetVol{999}{2026}

\newcommand{\TabCapp}[2]{\begin{center}\parbox[t]{#1}{\centerline{
  \small {\spaceskip 2pt plus 1pt minus 1pt T a b l e}
  \refstepcounter{table}\thetable}
  \vskip2mm
  \centerline{\footnotesize #2}}
  \vskip3mm
\end{center}}

\newcommand{\MakeTableee}[4]{\begin{table}[htb]\TabCapp{#2}{#3}
  \begin{center} \TableFont \begin{tabular}{#1} #4
  \end{tabular}\end{center}\end{table}}

\newfont{\bb}{ptmbi8t at 12pt}
\newfont{\bbb}{cmbxti10}
\newfont{\bbbb}{cmbxti10 at 9pt}

\begin{document}

\begin{Titlepage}
\Title{Thirty Circumbinary Disk Occultation Systems (KH~15D-like stars) from the OGLE Project\footnote{Based on observations obtained with the 1.3-m Warsaw telescope at the Las Campanas Observatory of the Carnegie Institution for Science and 3.58-m New Technology Telescope of the European Southern Observatory.}}

\Author{M.\,A.~~U~r~b~a~n~o~w~i~c~z$^1$,~~
I.~~S~o~s~z~y~\'n~s~k~i$^1$,~~
P.~~P~i~e~t~r~u~k~o~w~i~c~z$^1$,~~
A.~~U~d~a~l~s~k~i$^1$,\\
P.~~M~r~{\'o}~z$^1$,~~
M.~~W~r~o~n~a$^{1,2}$,~~
M.~~R~a~t~a~j~c~z~a~k$^1$,~~
M.\,K.~~S~z~y~m~a~{\'n}~s~k~i$^1$,\\
J.~~S~k~o~w~r~o~n$^1$,~~
D.\,M.~~S~k~o~w~r~o~n$^1$,~~
R.~~P~o~l~e~s~k~i$^1$,~~
S.~~K~o~z~{\l}~o~w~s~k~i$^1$,\\
P.~~I~w~a~n~e~k$^1$,~~
K.~~U~l~a~c~z~y~k$^{3,1}$,~~
K.~~R~y~b~i~c~k~i$^{1}$,~~
M.~~G~r~o~m~a~d~z~k~i$^1$,\\
and~~M.~~M~r~{\'o}~z$^1$
}
{$^1$Astronomical Observatory, University of Warsaw, Al.~Ujazdowskie~4, 00-478~Warszawa, Poland\\
$^2$Departament of Astrophysics and Planetary Sciences, Villanova University, 800 Lancaster
Ave., Villanova, PA19085, USA\\
$^3$Department of Physics, University of Warwick, Gibbet Hill Road, Coventry, CV4~7AL,~UK
}
\Received{May 999, 2026}
\end{Titlepage}

\vspace*{-2mm}
\Abstract{We present a catalog of 30 stars that are candidates for KH~15D-like binary systems, in which the observed brightness variations are caused by a circumbinary dusty disk that periodically obscures at least one of the stellar components as it moves along its orbit. Thanks to the regular observations conducted within the Optical Gravitational Lensing Experiment (OGLE) project, we provide unique light curves in the $I$ and $V$ bands with very long time baselines, in some cases beginning as early as 1997 and extending to the present day. Such long-term monitoring allows us to identify changes in eclipse widths, amplitudes, and light-curve shapes on timescales of many years. We highlight several circumbinary disk occultation (CBO) systems of particular interest and present spectra for three of them.}{Accretion, accretion disks, Eclipses, binaries: general, Stars: individual: KH~15D}

\section{Introduction}

In 1998, Kristin E. Kearns and William Herbst published the results of observations of the young cluster NGC 2264, during which they discovered, among other objects, a peculiar variable star designated KH~15D after the discoverers (Kearns \& Herbst 1998). KH 15D is composed of two pre-main sequence T~Tauri stars on an eccentric orbit ($e \approx 0.6$) and surrounded by a circumbinary dusty disk. The dust grains in the disk have sizes of up to $10\;\mu\mathrm{m}$ (Agol \etal 2004, Hamilton \etal 2005). This dusty disk has a radius of about 3~AU and is responsible for the observed brightness variations, producing broad and deep (up to 3~mag) eclipses with a period of 48.37~d when one of the stellar components becomes obscured by the disk as it moves along its orbit. Moreover, the disk itself undergoes precession, which results in long-term changes in the light-curve morphology. The archival observations of KH~15D obtained since 1951 (Chiang \& Murray-Clay 2004, Winn \etal 2004, Soto \etal 2020) revealed that in the past the precessing disk initially left both components of the system visible, then obscured one of them, and later for some time obscured both simultaneously, which manifested itself as a decrease in the maximum brightness of the system. More recently, the second component has begun to reappear, although the maximum brightness of the system now differs slightly from its level prior to 2005. Additionally, KH~15D exhibits intrinsic variability at the level of 0.1~mag, typical of spotted T~Tauri stars (Windemuth \& Herbst 2014, Aronow \etal 2018).

In the decades since the discovery of KH~15D, additional objects with similar properties have been reported (Plavchan \etal 2008, 2013, Rodr\'iguez-Ledesma \etal 2012, 2013, Zhu \etal 2022, Bernhard \etal 2024, Bernhard \& Lloyd 2024, Hu \etal 2024, 2026, Lucas \etal 2024). Stars of this type are commonly referred to as circumbinary disk occultation (CBO) systems. To date, 15 such objects have been identified, based on the compilation presented by Hu \etal (2026). Most of them are classified in the literature as young stellar objects (YSOs) and are associated with star-forming regions or young stellar clusters. Age estimates are available for some of these stars, typically in the range of $\sim 1$--$50~\mathrm{Myr}$ (Hamilton \& Herbst 2001, Plavchan \etal 2008, Rodr\'iguez-Ledesma \etal 2012, Hu \etal 2024, 2026). Their orbital periods span from several tens to hundreds of days, while the occultation durations range from about 20\% to over 80\% of the period.

The study of such objects provides valuable insight into the physical processes occurring in young binary systems, particularly enabling a deeper understanding of the dynamics and evolution of circumbinary disks. It also offers an opportunity to investigate planet formation in binary systems, where planets may form within such disks.

In this paper, we present a catalog of 30 variable stars whose properties are consistent with those of CBO systems. All objects were identified in the photometric databases of the Optical Gravitational Lensing Experiment (OGLE). The remainder of the paper is organized as follows. In Section 2, we describe the photometric and spectroscopic observations of our CBO systems. In Section~3, we outline the selection of CBO candidates from the OGLE databases. Section~4 presents our sample of CBO systems together with their long-term OGLE photometry. In Section~5, we discuss selected objects of particular interest and present spectra for three of our CBO systems. We summarize our results in Section~6.

\section{Observations and Data Reduction}

\subsection{Photometry}

The 30 CBO systems described here were observed with the 1.3-m Warsaw Telescope at Las Campanas Observatory, Chile, as part of the OGLE sky survey. Some of the observations presented in this paper extend back to 1997, \ie to the beginning of the second phase of the OGLE project (OGLE-II; Udalski \etal 1997), when the telescope was equipped with a CCD camera with a detector size of $2048 \times 2048$ pixels. The pixel size was $24~\mu\mathrm{m}$, corresponding to a scale of 0.417~arcsec/pixel. The observations covered small regions of the Galactic bulge and disk, as well as parts of the Small Magellanic Cloud and the Large Magellanic Cloud.

The OGLE-III project (Udalski \etal 2008) began in June 2001. During this phase, the surveyed sky area toward Galactic bulge and Magellanic Clouds was increased. The observations were carried out using a mosaic camera composed of eight CCD chips, each with dimensions of $2048 \times 4096$ pixels, giving a scale of 0.26~arcsec/pixel.

The most recent data come from the OGLE-IV survey, which started in March 2010. During this phase, the coverage of the observed fields was further expanded, and a mosaic CCD camera consisting of 32 chips with a total of 268.4 megapixels was used. Each chip has dimensions of $2048 \times 4102$ pixels, and the pixel scale is practically equal to that of OGLE-III, 0.26~arcsec/pixel. A detailed description of the instruments, photometric reductions, and astrometric calibrations of OGLE-IV observations is presented in Udalski \etal (2015).

All stars in our catalog were identified within an area of approximately 3000 square degrees covering the Galactic bulge and disk of the Milky Way. The OGLE observations are obtained in the {\it I} and {\it V} bands, calibrated to the standard Johnson-Cousins photometric system, with the vast majority of measurements collected in the {\it I} band. Observations in the {\it V} band constitute at most a few percent of the available photometric data and are utilized chiefly for color determination.

The time baseline, cadence, and number of data points in the OGLE light curves depend strongly on the position of a given star on the sky. Some objects observed toward the Galactic bulge have {\it I}-band light curves containing up to about $20\;000$ measurements obtained between 1997 and 2025, with a gap caused by the COVID-19 pandemic in 2020--2022. In contrast, the light curves of objects located in the Milky Way disk and in the outer regions of the Galactic bulge contain between 115 and 413 points and typically span about 10 years.

\subsection{Spectroscopy}
For three CBO systems reported in this paper, we obtained low-resolution spectroscopic observations. Objects OGLE-CBO-021 and OGLE-CBO-024 were observed on 2021 July 1/2, object OGLE-CBO-018 was observed on 2021 July 3/3, all of them as part of the ESO programme 105.20EF.001. The spectra were collected with the ESO Faint Object Spectrograph and Camera 2 (EFOSC2) mounted to the 3.58-m New Technology Telescope (NTT) located at La Silla Observatory, Chile. Details about this instrument can be found in Buzzoni \etal (1984). Thin cirrus clouds were present during the two nights. The spectra were taken with grism \#4 covering wavelengths 4085--7520~$\mathrm{\textup{\r{A}}}$ at the slit width of $1\zdot\arcs0$ and $2\times2$ binning readout, which provided a spectral resolution of about 11~$\mathrm{\textup{\r{A}}}$ at 5000~$\mathrm{\textup{\r{A}}}$. In the case of CBO objects, we collected two consecutive exposures to remove possible cosmic rays in the spectra. The exposure times were calculated with the help of the ESO Exposure Time Calculator\footnote{https://www.eso.org/observing/etc/}. For the wavelength calibrations, we used the He-Ar lamp. For the flux calibrations, three spectrophotometric standard stars were selected. Bias and dome flat-field images were taken at dawn. We reduced the spectra using the IRAF package\footnote{IRAF was distributed by the National Optical Astronomy Observatory, USA, which is operated by the Association of Universities for Research in Astronomy, Inc., under a cooperative agreement with the National Science Foundation.} (Tody 1986, 1993). Debiasing, flat-fielding, wavelength and flux calibrations were performed in the standard way. Log of spectroscopic observations in presented in Table 1.

\MakeTableee{l@{\hspace{15pt}}
c@{\hspace{20pt}}   
c@{\hspace{5pt}}
c@{\hspace{5pt}}
c@{\hspace{5pt}}
c@{\hspace{5pt}}}{12.5cm}
{Exposure Times and Dates of Spectroscopic Observations}
{\hline
\noalign{\vskip3pt}
Object    &    $t_{\mathrm{exp1}}$ & $t_{\mathrm{exp2}}$ &  $\mathrm{exp1_{start}}$ &  $\mathrm{exp2_{start}}$ \\
          &  [s]  &    [s]    &     JD & JD     \\
\noalign{\vskip3pt}
\hline
\noalign{\vskip3pt}
OGLE-CBO-018 &  2400 &   2400  &  2459399.640 &  2459399.670 \\
OGLE-CBO-021 &  3000 &   3000  &  2459397.712 &  2459397.747 \\
OGLE-CBO-024 &  3000 &   1887 &   2459397.605 &  2459397.640 \\
\noalign{\vskip3pt}
\hline}

\MakeTableee{l@{\hspace{12pt}}
c@{\hspace{12pt}}   
c@{\hspace{8pt}}
c@{\hspace{8pt}}
c@{\hspace{8pt}}
c@{\hspace{8pt}}
c@{\hspace{2pt}}}{12.0cm}
{Catalog of CBO System Candidates}
{\hline
\noalign{\vskip3pt}
Name & Ra & Dec & Alternative name & Ref. \\
\noalign{\vskip3pt}
\hline
\noalign{\vskip3pt}
OGLE-CBO-001 & $06\uph38\upm26\zdot\ups76$ & $\hphantom{-}04\arcd38\arcm10\zdot\arcs5$ &  &  \\ 

OGLE-CBO-002 & $06\uph40\upm06\zdot\ups57$ & $-05\arcd51\arcm23\zdot\arcs7$ &  &  \\ 

OGLE-CBO-003 & $07\uph04\upm12\zdot\ups92$ & $-11\arcd24\arcm03\zdot\arcs2$ & ZTF J070412.91-112403.2 & (1) \\ 

OGLE-CBO-004 & $07\uph14\upm45\zdot\ups40$ & $-09\arcd01\arcm52\zdot\arcs2$ & Bernhard-2 & (2,3) \\ 

OGLE-CBO-005 & $08\uph14\upm59\zdot\ups00$ & $-35\arcd42\arcm21\zdot\arcs9$ &  &  \\ 

OGLE-CBO-006 & $08\uph41\upm51\zdot\ups77$ & $-42\arcd20\arcm24\zdot\arcs6$ &  &  \\ 

OGLE-CBO-007 & $08\uph49\upm41\zdot\ups60$ & $-44\arcd21\arcm20\zdot\arcs0$ &  &  \\ 

OGLE-CBO-008 & $09\uph41\upm26\zdot\ups96$ & $-47\arcd57\arcm34\zdot\arcs3$ &  &  \\ 

OGLE-CBO-009 & $10\uph39\upm16\zdot\ups11$ & $-58\arcd56\arcm34\zdot\arcs5$ &  &  \\ 

OGLE-CBO-010 & $10\uph57\upm35\zdot\ups17$ & $-60\arcd10\arcm17\zdot\arcs9$ &  &  \\ 

OGLE-CBO-011 & $11\uph11\upm29\zdot\ups78$ & $-61\arcd12\arcm45\zdot\arcs5$ &  &  \\ 

OGLE-CBO-012 & $11\uph18\upm46\zdot\ups38$ & $-56\arcd12\arcm40\zdot\arcs8$ &  &  \\ 

OGLE-CBO-013 & $11\uph24\upm02\zdot\ups41$ & $-60\arcd16\arcm38\zdot\arcs0$ & GDS\_J1124024-601637 & (4) \\ 

OGLE-CBO-014 & $13\uph18\upm10\zdot\ups69$ & $-62\arcd38\arcm41\zdot\arcs6$ &  &  \\ 

OGLE-CBO-015 & $14\uph35\upm41\zdot\ups82$ & $-60\arcd00\arcm04\zdot\arcs6$ &  &  \\ 

OGLE-CBO-016 & $15\uph25\upm11\zdot\ups59$ & $-59\arcd09\arcm17\zdot\arcs0$ &  &  \\ 

OGLE-CBO-017 & $16\uph10\upm11\zdot\ups25$ & $-51\arcd00\arcm28\zdot\arcs0$ &  &  \\ 

OGLE-CBO-018 & $16\uph50\upm57\zdot\ups01$ & $-41\arcd11\arcm26\zdot\arcs0$ &  &  \\ 

OGLE-CBO-019 & $16\uph58\upm05\zdot\ups89$ & $-39\arcd18\arcm41\zdot\arcs3$ &  &  \\ 

OGLE-CBO-020 & $17\uph48\upm36\zdot\ups26$ & $-35\arcd37\arcm12\zdot\arcs0$ & OGLE-BLG-LPV-046008 & (5) \\ 

OGLE-CBO-021 & $17\uph52\upm23\zdot\ups06$ & $-29\arcd33\arcm50\zdot\arcs8$ &  &  \\ 

OGLE-CBO-022 & $17\uph55\upm36\zdot\ups25$ & $-29\arcd09\arcm22\zdot\arcs5$ &  &  \\ 

OGLE-CBO-023 & $17\uph59\upm38\zdot\ups45$ & $-29\arcd33\arcm21\zdot\arcs9$ & V5875 Sgr & (5,6) \\ 

OGLE-CBO-024 & $18\uph01\upm07\zdot\ups87$ & $-30\arcd19\arcm32\zdot\arcs2$ &  	 OGLE BUL-SC38 V0489 & (7,8) \\ 

OGLE-CBO-025 & $18\uph04\upm29\zdot\ups02$ & $-26\arcd15\arcm39\zdot\arcs10$ &  &  \\ 

OGLE-CBO-026 & $18\uph05\upm02\zdot\ups30$ & $-24\arcd25\arcm01\zdot\arcs2$ & VVV J180502.29-242501.4 & (9) \\ 

OGLE-CBO-027 & $18\uph07\upm26\zdot\ups39$ & $-26\arcd56\arcm36\zdot\arcs60$ & OGLE-BLG-LPV-213411 & (5) \\ 

OGLE-CBO-028 & $18\uph16\upm27\zdot\ups04$ & $-19\arcd47\arcm21\zdot\arcs10$ & ZTF-CBO-2 & (10) \\ 

OGLE-CBO-029 & $18\uph16\upm39\zdot\ups95$ & $-15\arcd44\arcm33\zdot\arcs90$ & ZTF-CBO-3 & (10) \\ 

OGLE-CBO-030 & $18\uph35\upm26\zdot\ups57$ & $-12\arcd51\arcm40\zdot\arcs30$ & ZTF-CBO-5 & (10) \\ 

\noalign{\vskip3pt}
\hline
\noalign{\vskip5pt}
\multicolumn{5}{p{12cm}}{(1) (Bernhard \etal 2024), (2) (Zhu \etal 2022), (3) (Hu \etal 2024), (4) (Hackstein \etal 2015), (5) (Soszy\'nski \etal 2013), (6) (Bernhard \etal 2013), (7) (Groenewegen \& Blommaert 2005), (8) (Matsunaga \etal 2005), (9) Lucas \etal 2024, (10) (Hu \etal 2026)}
\label{tab:tab1}
}

\MakeTableee{l@{\hspace{12pt}}
c@{\hspace{12pt}}   
c@{\hspace{8pt}}
c@{\hspace{8pt}}
c@{\hspace{8pt}}
c@{\hspace{8pt}}
c@{\hspace{2pt}}}{10.0cm}
{Selected Observational Parameters of CBO Stars in the OGLE Collection}
{\hline
\noalign{\vskip3pt}
Name         & $P$                & $I_{\mathrm{max}}$ & $V_{\mathrm{max}}$   & $A_I$ & Occ. W.  & $T_{\mathrm{O}}$   \\
\hphantom{0} & $\mathrm{[d]}$ & $\mathrm{[mag]}$ & $\mathrm{[mag]}$ & $\mathrm{[mag]}$ & $\%$ & $\mathrm{[HJD]}$\\
\noalign{\vskip3pt}
\hline
\noalign{\vskip3pt}
OGLE-CBO-001 &$\hphantom{0}34.833$ & $16.16$ & $17.54$ & 2.06 & 29 & $2\,460\,012.337$ \\
OGLE-CBO-002 &$\hphantom{0}80.731$ & $14.57$ & $15.68$ & 0.18 & 54 & $2\,460\,045.415$ \\
OGLE-CBO-003 &$\hphantom{0}34.578$ & $15.70$ & $17.61$ & 0.94 & 39 & $2\,460\,002.639$ \\
OGLE-CBO-004 &$\hphantom{0}63.297$ & $15.98$ & $17.44$ & 1.75 & 47 & $2\,460\,056.346$ \\
OGLE-CBO-005 &$\hphantom{0}65.189$ & $17.05$ & $18.87$ & 2.18 & 32--64 & $2\,460\,015.248$ \\
OGLE-CBO-006 &$\hphantom{0}42.263$ & $16.47$ & ---     & 0.79 & 19 & $2\,460\,028.127$ \\
OGLE-CBO-007 &$\hphantom{0}80.681$ & $18.32$ & $21.01$ & 0.60 & 52 & $2\,460\,068.401$ \\
OGLE-CBO-008 &           $759.290$ & $15.78$ & $18.66$ & 0.80 & 82 & $2\,460\,545.660$ \\
OGLE-CBO-009 &           $280.957$ & $16.93$ & $18.34$ & 1.71 & 77 & $2\,460\,262.273$ \\
OGLE-CBO-010 &$\hphantom{0}48.228$ & $18.94$ & $20.67$ & 2.39 & 0--51 & $2\,458\,591.924$ \\
OGLE-CBO-011 &$\hphantom{0}97.005$ & $15.79$ & $17.36$ & 0.44 & 43 & $2\,460\,091.958$ \\
OGLE-CBO-012 &$\hphantom{0}43.274$ & $13.07$ & $13.51$ & 0.14 & 66 & $2\,460\,032.535$ \\
OGLE-CBO-013 &           $155.301$ & $13.47$ & $14.05$ & 0.69 & 20 & $2\,460\,126.421$ \\
OGLE-CBO-014 &           $107.086$ & $16.48$ & $18.49$ & 0.69 & 73 & $2\,460\,000.922$ \\
OGLE-CBO-015 &$\hphantom{0}64.686$ & $16.74$ & $19.32$ & 0.46 & 68 & $2\,460\,057.022$ \\
OGLE-CBO-016 &           $121.601$ & $16.16$ & ---     & 1.11 & 26 & $2\,460\,110.831$ \\
OGLE-CBO-017 &$\hphantom{0}73.674$ & $16.84$ & $19.33$ & 3.17 & 27--49 & $2\,460\,045.012$ \\
OGLE-CBO-018 &$\hphantom{0}51.376$ & $14.39$ & $17.91$  & 3.38 & 87--92 & $2\,460\,006.748$ \\
OGLE-CBO-019 &$\hphantom{0}61.052$ & $16.83$ & $18.27$ & 2.53 & 20 & $2\,460\,002.153$ \\
OGLE-CBO-020 &           $354.239$ & $14.47$ & $15.33$ & 0.69 & 52 & $2\,460\,308.345$ \\
OGLE-CBO-021 &           $141.751$ & $18.76$ & $20.75$ & 1.57 & 27 & $2\,460\,023.327$ \\
OGLE-CBO-022 &           $227.640$ & $15.61$ & $17.71$ & 0.06 & 79 & $2\,460\,074.835$ \\
OGLE-CBO-023 &           $431.156$ & $13.83$ & $15.28$ & 3.18 & 57--92 & $2\,460\,052.736$ \\
OGLE-CBO-024 &           $272.119$ & $15.23$ & $18.78$ & 2.64 & 78 & $2\,460\,269.827$ \\
OGLE-CBO-025 &$\hphantom{0}18.181$ & $17.97$ & $20.05$ & 2.21 & 56--68 & $2\,458\,256.704$ \\
OGLE-CBO-026 &$\hphantom{0}59.270$ & $15.82$ & $18.83$ & 2.76 & 29--91 & $2\,458\,774.855$ \\
OGLE-CBO-027 &           $259.675$ & $14.50$ & $18.14$ & 1.47 & 56--73 & $2\,460\,025.823$ \\
OGLE-CBO-028 &$\hphantom{0}74.337$ & $16.43$ & ---     & 2.81 & 54 & $2\,460\,073.748$ \\
OGLE-CBO-029 &           $151.017$ & $15.58$ & $16.65$ & 1.43 & 42 & $2\,460\,029.690$ \\
OGLE-CBO-030 &$\hphantom{0}61.282$ & $16.72$ & $18.53$ & 0.97 & 43 & $2\,460\,025.179$ \\
\noalign{\vskip3pt}
\hline
\noalign{\vskip5pt}
\multicolumn{7}{p{10.0cm}}{$P$ - variability period; $I_{\mathrm{max}}$ - $I$-band maximal brightness level; $V_{\mathrm{max}}$ - $V$-band maximal brightness level; $A_I$ - $I$-band amplitude; Occ. W. - occultation width expressed in variability period percentage;  $T_{\mathrm{O}}$ - occultation epoch.}
}

\section{Selection of KH~15D-like Stars}

The objects in our catalog were identified in the OGLE photometric databases during searches for other classes of variable stars, including long-period variables (Soszy\'nski \etal 2013), dwarf novae (Mr\'oz \etal 2013), eclipsing binaries (Piet\-ru\-ko\-wicz \etal 2013, Soszy\'nski \etal 2016), Cepheids (Udalski \etal 2018), and eccentric ellipsoidal variables (AKA heartbeat stars; Wrona \etal 2022). Visual inspection of the light curves revealed unusual variables that spend most of their time in one of two distinct brightness states, within which the flux remains approximately constant, while transitions between these states occur on relatively short timescales. The resulting light curves resemble those of eclipsing systems, but most of them exhibit broad, flat-bottomed minima that, in extreme cases, persist for up to 80\% of the orbital period. The amplitudes of the variability span a wide range, from a few hundredths of a magnitude to several magnitudes.

Such light-curve shapes are characteristic of CBO systems, such as KH 15D. We therefore classify these stars as candidates for this class of objects, however, definitive confirmation of their nature will require spectroscopic follow-up observations. To increase the completeness of our catalog, we supplemented the sample with five additional CBO systems (Hu \etal 2026) whose photometry is available in the OGLE databases but which had not previously been identified as particularly distinctive variable stars. Finally, the OGLE collection of KH~15D-like stars comprises 30 objects.

\section{OGLE Collection of CBO Systems}

The OGLE sample of CBO systems is summarized in Tables~2 and~3. The stars are ordered by increasing right ascension and assigned identifiers of the form OGLE-CBO-NNN, where NNN is a sequential number. In Table~2, we list the equatorial coordinates (J2000.0) of the objects, together with their alternative designations from the International Variable Star Index (VSX; Watson \etal 2006). Only six of these stars have previously been reported as CBO systems (designated E-DO in the VSX nomenclature), whereas the remaining objects were cataloged as variable stars of other types, among others, as long-period variables or cataclysmic variables.

Table~3 lists the parameters of the CBO systems derived from the OGLE photometry: orbital periods, maximum brightness in the {\it I}- and {\it V}-band filters, {\it I}-band amplitudes, occultation-to-period ratios, and epochs of occultation. The electronic versions of Tables~2 and~3, together with the OGLE time-series photometry, are available via the OGLE Internet Archive.

\begin{center}
{\it https://www.astrouw.edu.pl/ogle/ogle4/OCVS/CBO/}
\end{center}

Figs.~1 and~2 present the phase-folded {\it I}-band light curves of all CBO systems in our catalog. In some cases, the light-curve morphology varies, exhibiting shifts in the ingress and egress times as well as changes in eclipse depth, including the complete disappearance of occultations. Selected examples of noteworthy behavior among CBO systems are discussed in Section~5.

\MakeTableee{l@{\hspace{15pt}}
c@{\hspace{20pt}}   
c@{\hspace{5pt}}
c@{\hspace{5pt}}}{12.5cm}
{CBO systems positionally and kinematically coincident with star clusters}
{\hline
\noalign{\vskip3pt}
Name & Cluster Name & Age $\mathrm{[Myr]}$ \\
\noalign{\vskip3pt}
\hline
\noalign{\vskip3pt}
OGLE-CBO-001  &  Collinder 107    & \hphantom{0}14.6  \\
OGLE-CBO-003  &  vdBergh 92       & \hphantom{0}22.9  \\
OGLE-CBO-007  &  vdBergh-Hagen 54 & \hphantom{0}31.6  \\
OGLE-CBO-009  &  Bochum 10        & \hphantom{0}18.1  \\
OGLE-CBO-011  &  ASCC 65          & \hphantom{00}7.1  \\
OGLE-CBO-014  &  DBSB 85          & \hphantom{0}28.2  \\
OGLE-CBO-017  &  UBC 1545         & \hphantom{0}59.2  \\
OGLE-CBO-021  &  Ruprecht 134     & \hphantom{0}75.9  \\
OGLE-CBO-026  &  NGC 6530         & \hphantom{00}4.7  \\
\noalign{\vskip3pt}
\hline}

\begin{figure}[p]
\includegraphics[width=1.0\textwidth]{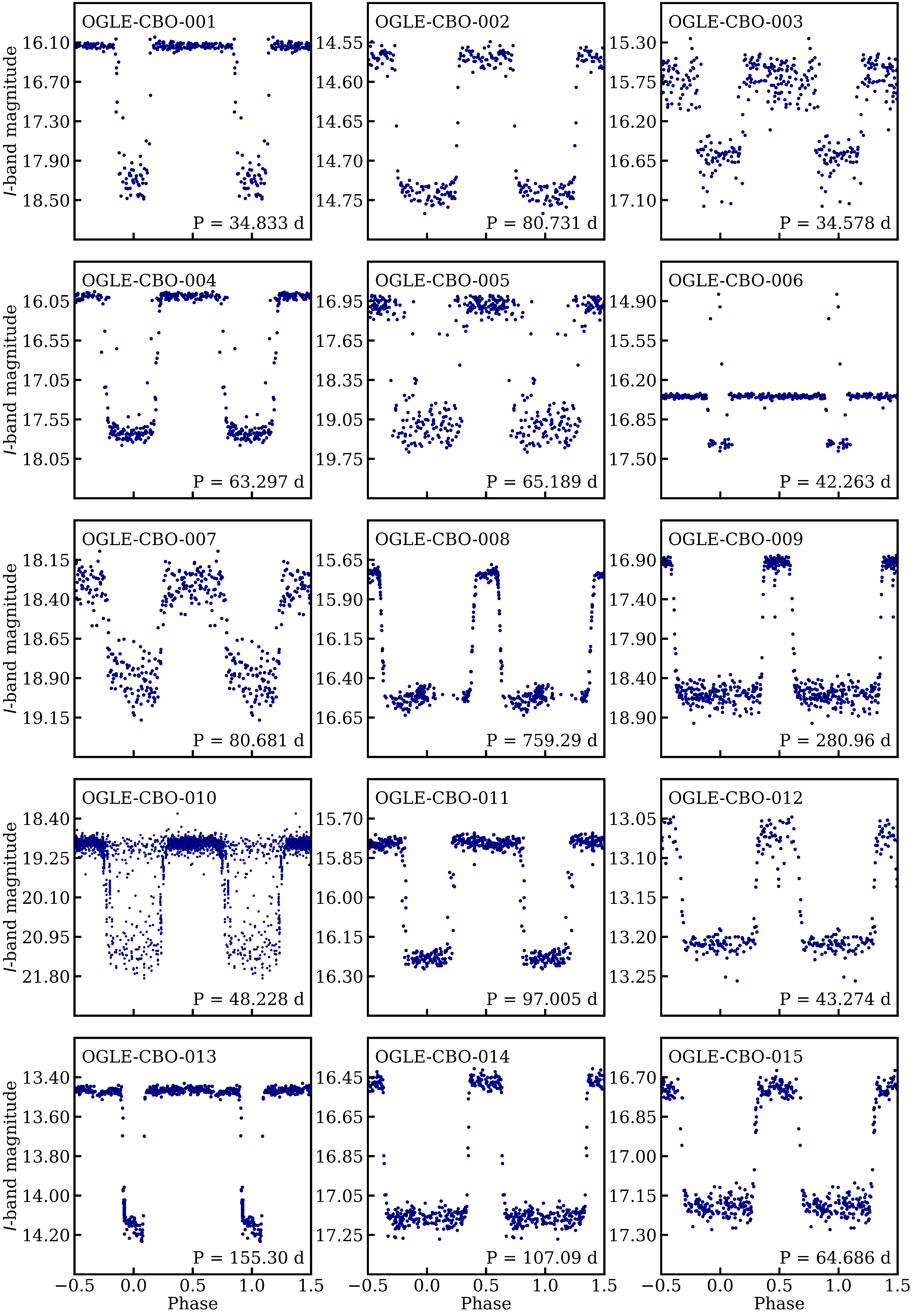}
\FigCap{Phase-folded {\it I}-band light curves of CBO systems from OGLE-CBO-001 to OGLE-CBO-015.}
\label{fig:fig1}
\end{figure}

\begin{figure}[p]
\includegraphics[width=1.0\textwidth]{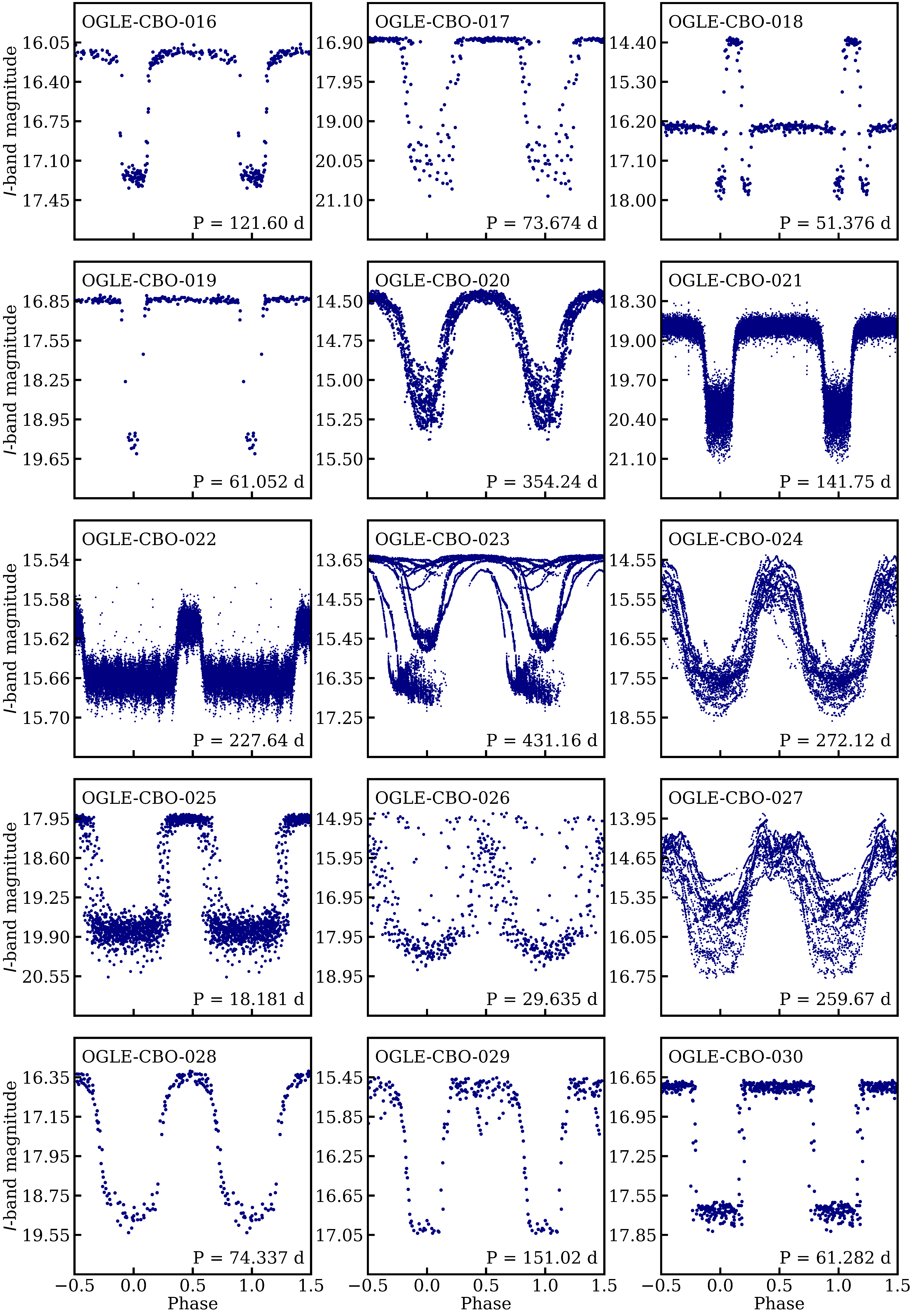}
\FigCap{Phase-folded {\it I}-band light curves of CBO systems from OGLE-CBO-016 to OGLE-CBO-030.}
\label{fig:fig2}
\end{figure}

Fig.~3 displays the on-sky distribution of the OGLE CBO systems in Galactic coordinates. The stars are concentrated toward the Galactic plane, as expected for very young objects. Of the 30 objects in our catalog, 28 are located within 4~degrees from the Galactic equator. We cross-matched our sample of CBO systems with star cluster catalogs published by Kharchenko et al. (2013) and Hunt and Reffert (2023). We find that nine systems lie within one cluster radius of the cluster center on the sky, and that their proper motions (Gaia Collaboration \etal 2023) are consistent with those of the clusters, indicating likely membership. Table~4 lists these objects together with the cluster ages reported by Kharchenko et al. (2013) and Hunt \& Reffert (2023). Note that all these clusters are young, with ages below 100~Myr.

\begin{figure}[t]
\includegraphics[width=1.0\textwidth]{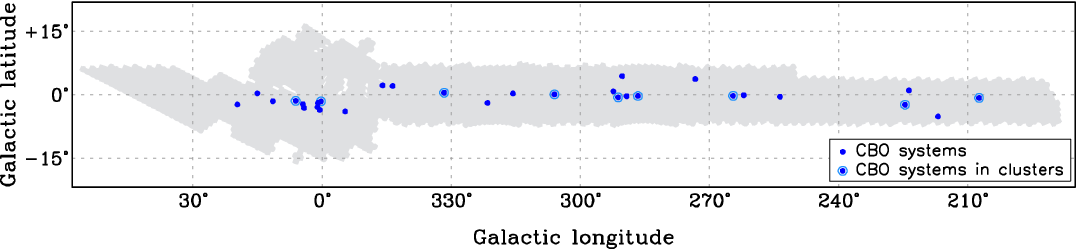}
\FigCap{On-sky distribution of CBO systems in Galactic coordinates. Points enclosed by light-blue circles indicate systems that are positionally coincident with star clusters listed in the catalogs of Kharchenko et al. (2013) and Hunt and Reffert (2023). The gray area shows the OGLE footprint in the Galactic bulge and disk.}
\label{fig:fig3}
\end{figure}

\section{Properties of Selected Objects}
In this section, we present CBO systems of particular interest. Many of the intriguing phenomena observed in these variables have been revealed thanks to the long-term, systematic photometric monitoring conducted by the OGLE survey. The extended temporal baseline of the OGLE data provides a robust foundation for a deeper understanding of the unique objects presented in this work and for constraining their nature.

\begin{figure}[b]
\includegraphics[width=1.0\textwidth]{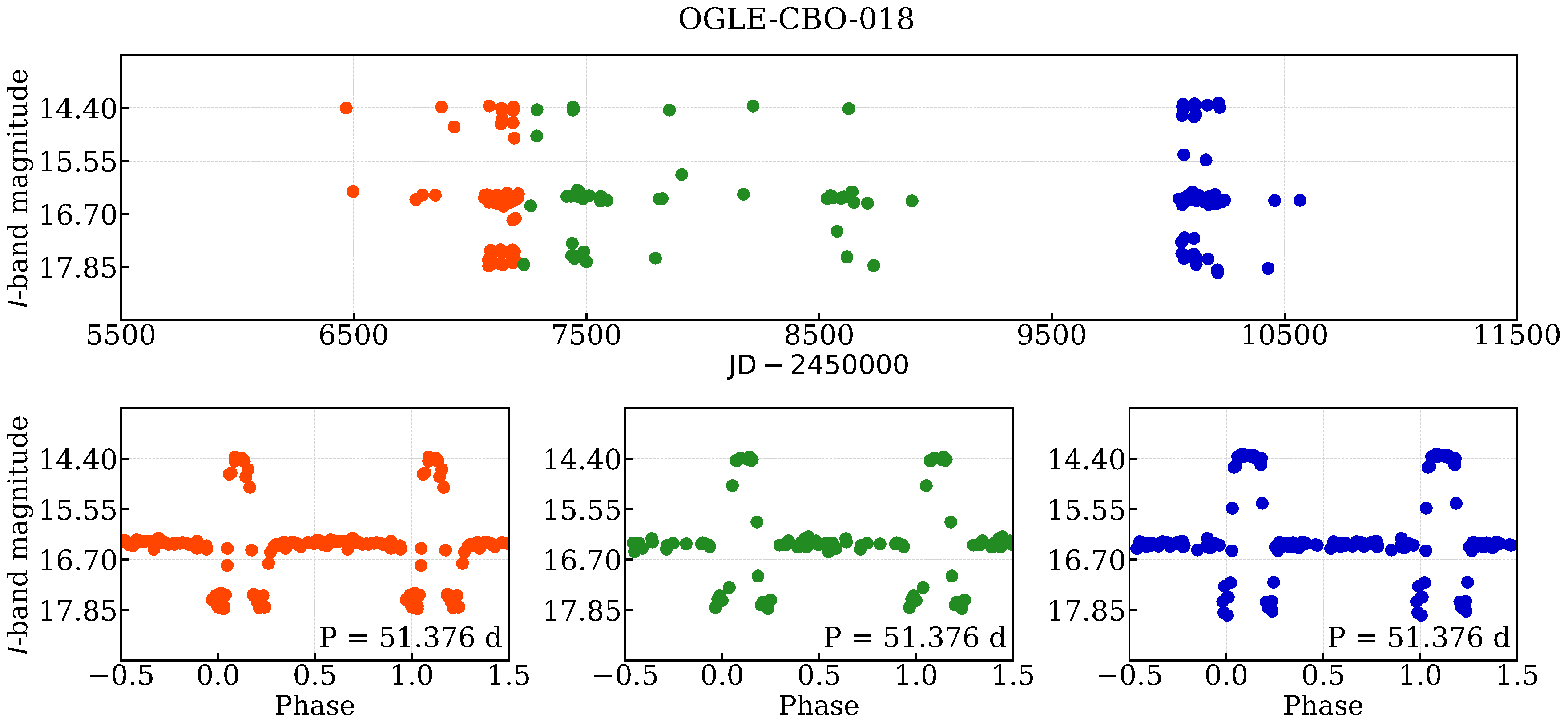}
\FigCap{{\it I}-band light curve of OGLE-CBO-018. The upper panel shows the time-series photometry obtained by the OGLE project during 2013--2024. The lower panels present phase-folded segments of this light curve for the intervals 2013--2015, 2016--2020, and 2023--2024. Each segment is shown in a different color.}
\label{fig:fig4}
\end{figure}

\subsection{OGLE-CBO-018}
OGLE-CBO-018 stands out among the stars in our catalog because of its unusual light curve (Fig.~4). For most of the cycle, the star remains at an approximately constant brightness level of $I \approx 16.3$~mag, although the light curve shows slight curvature. This phase is followed by a sudden decline in brightness of about 1.3~mag. The resulting minimum is brief and is followed by a rapid brightening of $\sim$3.2~mag to a maximum, during which the brightness again remains nearly constant for a short interval. Subsequently, a second minimum is observed, with a depth very similar to that of the first. Thus, OGLE-CBO-018 exhibits three distinct brightness levels over the course of its variability cycle, in contrast to most other CBO systems, which display only two states.

We include this object in the OGLE catalog of CBO systems because KH~15D itself has exhibited similar variability in the past. In particular, the modeled light curve of KH~15D presented by Poon \etal\ (2021) shows a striking resemblance to that of OGLE-CBO-018 during the 1995--1996 interval, with two minima and a single bright maximum exceeding the mean brightness level.

The lower panels of Fig.~4 present phase-folded light curves of OGLE-CBO-018 from the beginning, middle, and end of the OGLE observing baseline. Both minima become progressively narrower with time, while the duration of the maximum brightness increases. It is therefore plausible that these narrow minima will eventually disappear, and OGLE-CBO-018 will evolve toward a light curve morphology similar to that of other CBO systems, characterized by only two brightness levels.

\begin{figure}[b]
\includegraphics[width=1.0\textwidth]{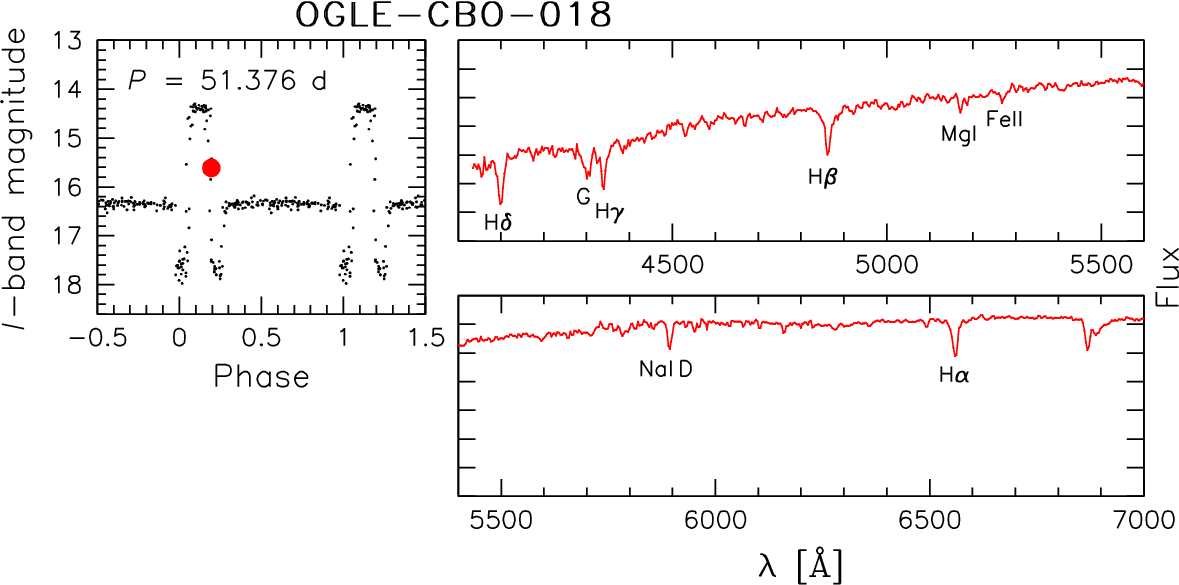}
\FigCap{Low-resolution spectra of object OGLE-CBO-018. Moments of the data acquision are marked with large red
dots in the phase-folded light curves (on the left).}
\label{fig:fig5}
\end{figure}

In Fig.~5, we present low-resolution spectrum of OGLE-CBO-018, obtained during the decline in brightness from maximum to minimum. Spectroscopic observations obtained for this object show that OGLE-CBO-018 is a G-type star with several metallic features (including the G band) absorption. No emission features are present in the spectrum.

\begin{figure}[h]
\includegraphics[width=1.0\textwidth]{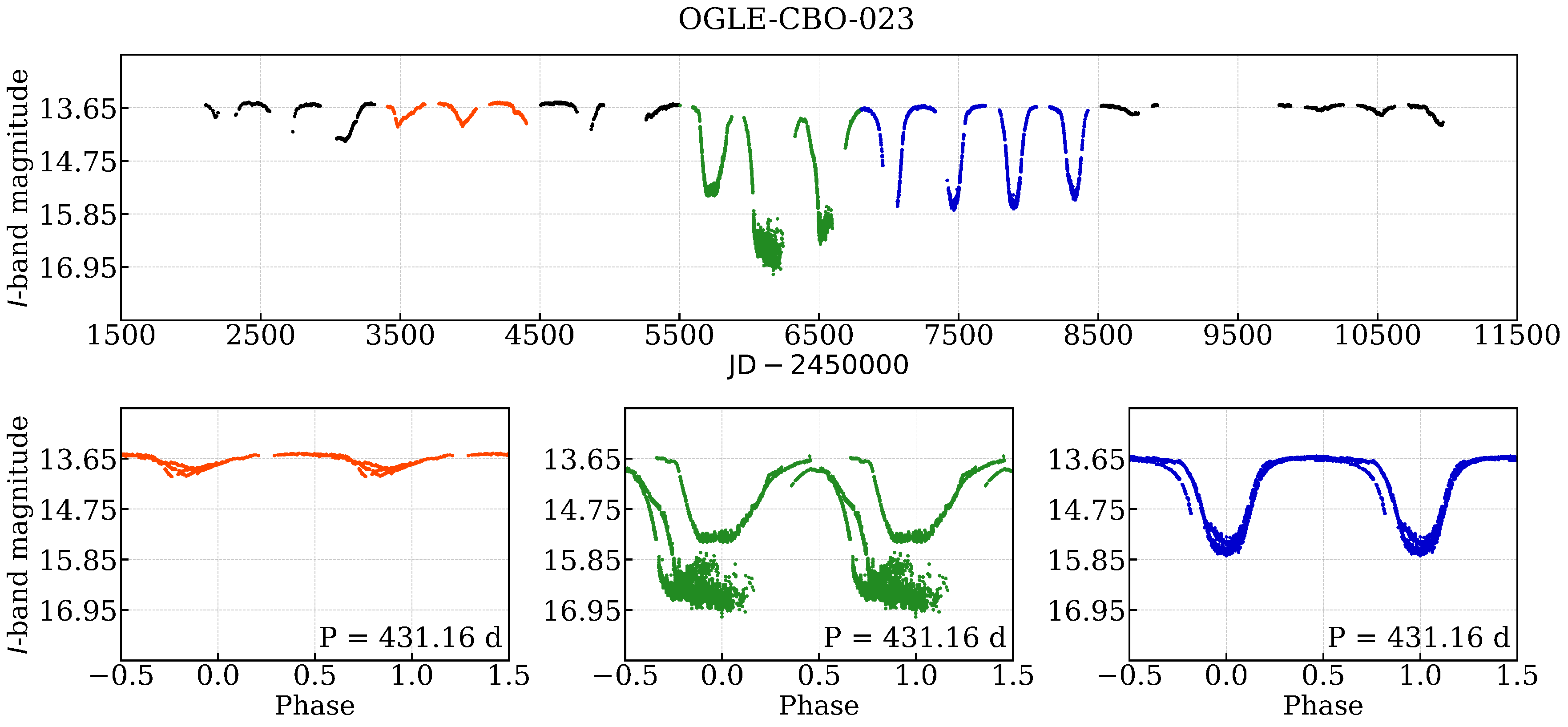}
\FigCap{{\it I}-band light curve of \textit{OGLE-CBO-023}. The upper panel shows unfolded time series. The lower three panels present phase-folded fragments of the light curve from the upper panel. Each fragment is marked with a different color.}
\label{fig:fig6}
\end{figure}

\begin{figure}[b]
\includegraphics[width=1.0\textwidth]{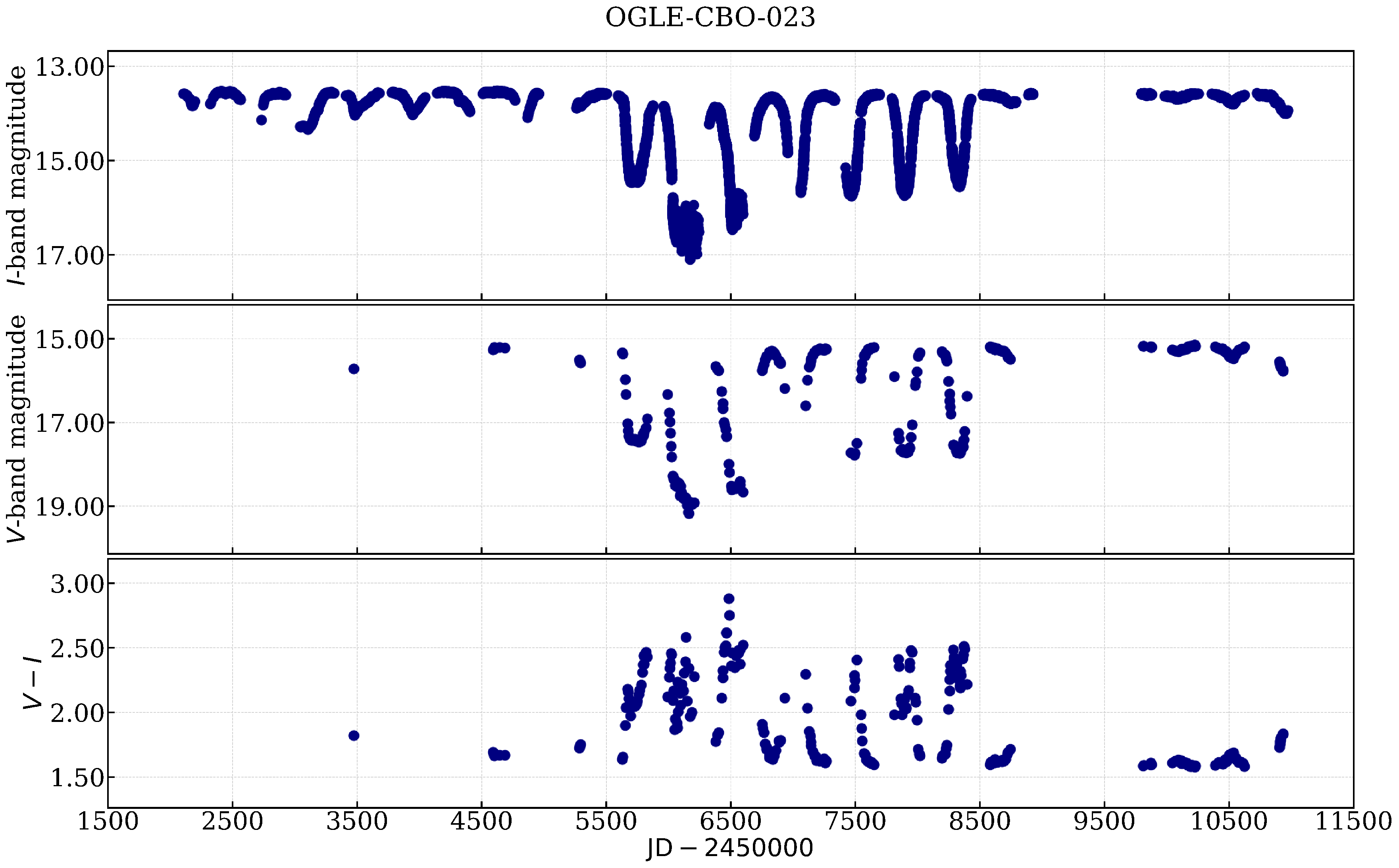}
\FigCap{{\it I}-band light curve (upper panel), {\it V}-band light curve (middle panel), and the $V-I$ color curve (lower panel) of \textit{OGLE-CBO-023}. Note that the star becomes redder during brightness declines.}
\label{fig:fig7}
\end{figure}

\subsection{OGLE-CBO-023}

The light curve of OGLE-CBO-023 (Fig.~6) has exhibited pronounced changes over time. During 2001--2010, the star showed relatively shallow eclipses, and the {\it I}-band light curve was asymmetric, with sharp minima (lower-left panel of Fig.~6). The morphology of the light curve changed rapidly in 2011, when its amplitude increased significantly, reaching about 3~mag in 2012 (lower-middle panel of Fig.~6). The star then entered a stable phase, during which the light curve became symmetric with minima that resembled broad eclipses (lower-right panel of Fig.~6). The most recent observations indicate that the star has returned to a state of low-amplitude variability. This behavior motivated us to include this object in the OGLE catalog of CBO systems. The brightness variations of OGLE-CBO-023 may in some sense resemble those of R CrB-type variables; however, due to the relatively strong periodicity of the brightness declines, we rule out the possibility that it is an R CrB-type variable.

In Fig.~7, we present the {\it I}-band light curve (upper panel), {\it V}-band light curve (middle panel) and the $V-I$ color curve (bottom panel) of OGLE-CBO-023. The amplitudes in the {\it V} band are larger than in the {\it I} band, \ie the star becomes redder at minimum brightness. Such behavior may indicate that the brightness drops are a result of obscuration by a thick dusty disk. 

\begin{figure}[b]
\includegraphics[width=1.0\textwidth]{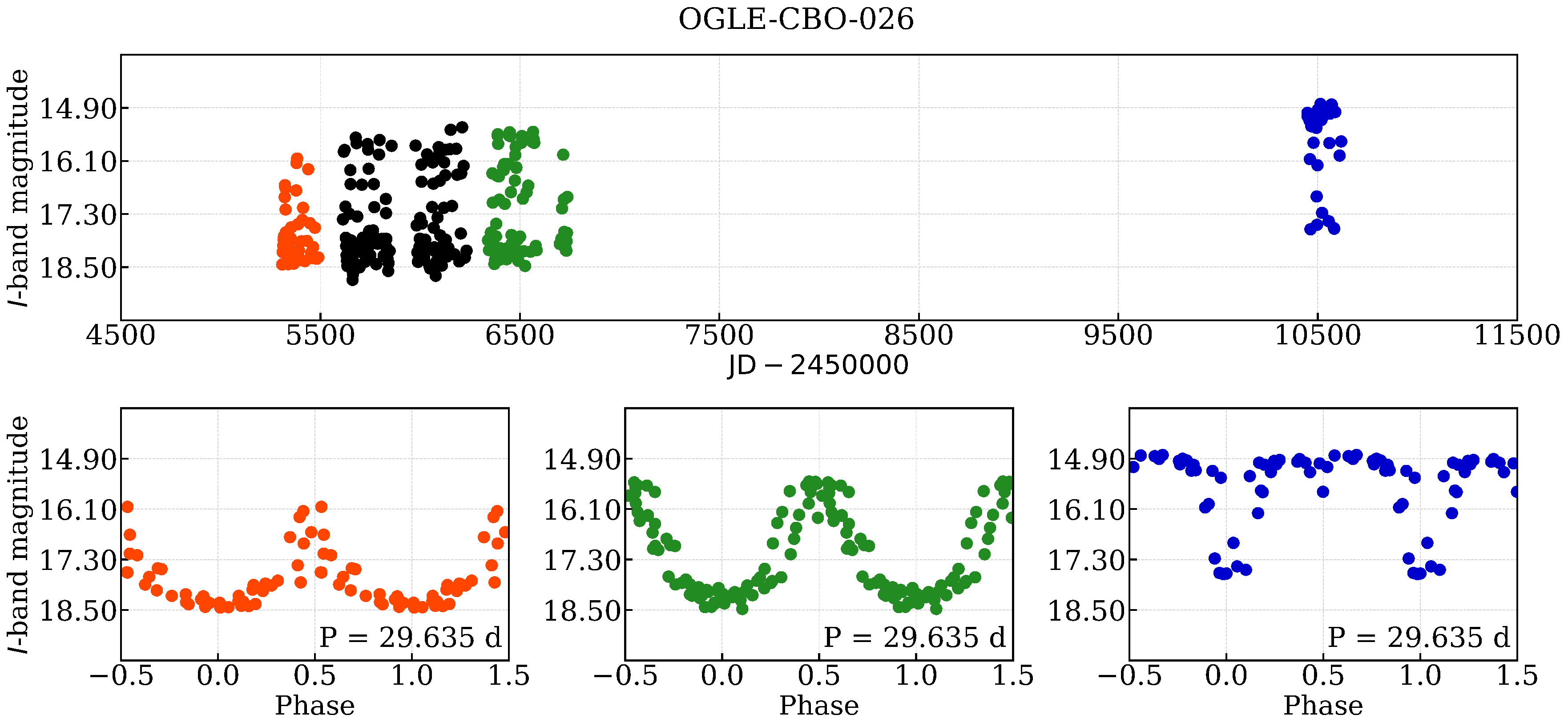}
\FigCap{{\it I}-band light curve of \textit{OGLE-CBO-026}. The upper panel shows unfolded time series. The lower three panels present phase-folded fragments of the light curve from the upper panel. Each fragment is marked with a different color.}
\label{fig:fig8}
\end{figure}

\subsection{OGLE-CBO-026}
OGLE-CBO-026 is a star exhibiting the largest changes in occultation width, ranging from approximately 29\% to 89\% of the orbital period. This object was initially classified as a dwarf nova by Mr{\'o}z \etal (2015), as the shape of its light curve resembled that class of variable stars. The maximum-brightness phase was short, while the minimum did not maintain a constant level. Unfortunately, the OGLE observations of this star contain a long, approximately ten-year gap between March 2014 and May 2024. After this interval, the light curve exhibits a markedly different morphology, with the maximum becoming much broader than the minimum. The evolution of the occultation width is illustrated in Fig.~8.

This star was previously described by Lucas \etal (2024) as a YSO exhibiting variability similar to that of KH~15D, with the distinction that, in their interpretation, both components of the system are alternately eclipsed by the disk. When one component is visible, the other is obscured by the disk, and after half of the period the situation reverses, with the second component becoming visible while the first is hidden. In their study, the variability period is reported as $59.35~\mathrm{d}$, which is approximately twice the value derived in this work. In our view, it is difficult to determine which period is correct, as the periodic variability in this object is superimposed on irregular brightness changes. Therefore, this issue warrants further investigation.

\subsection{CBO Systems with Variable Occultation Durations}
At least half of the CBO systems in our catalog exhibit variations in eclipse width, sometimes accompanied by changes in the amplitude of the brightness variations. The light curves of six representative objects of this type are shown in Figs.~9 and 10. As in the case of KH~15D, this effect may be attributed to the precession of a disk surrounding the binary system. As a result, the disk obscures one of the stellar components at different orbital phases over time. In the case of OGLE-CBO-010 (lower panel of Fig.~9), both the occultation width and amplitude gradually decreased until the eclipses disappeared completely. It is unclear whether the occultations have vanished temporarily or permanently. If a stable precessing disk surrounds this system, the eclipses should eventually reappear.

OGLE-CBO-027 (lower panel of Fig.~10) stands out from most objects in our catalogue due to its irregular brightness variations. Its brightness declines approximately linearly after reaching maximum, followed by a steeper decline. This behavior may be explained by the presence of a disk that does not have a sharp edge, as in KH~15D, and instead has a more inhomogeneous structure. In addition to changes in eclipse width and amplitude, we also observe variations in the mean brightness level.

\begin{figure}[p]
\includegraphics[width=1.0\textwidth]{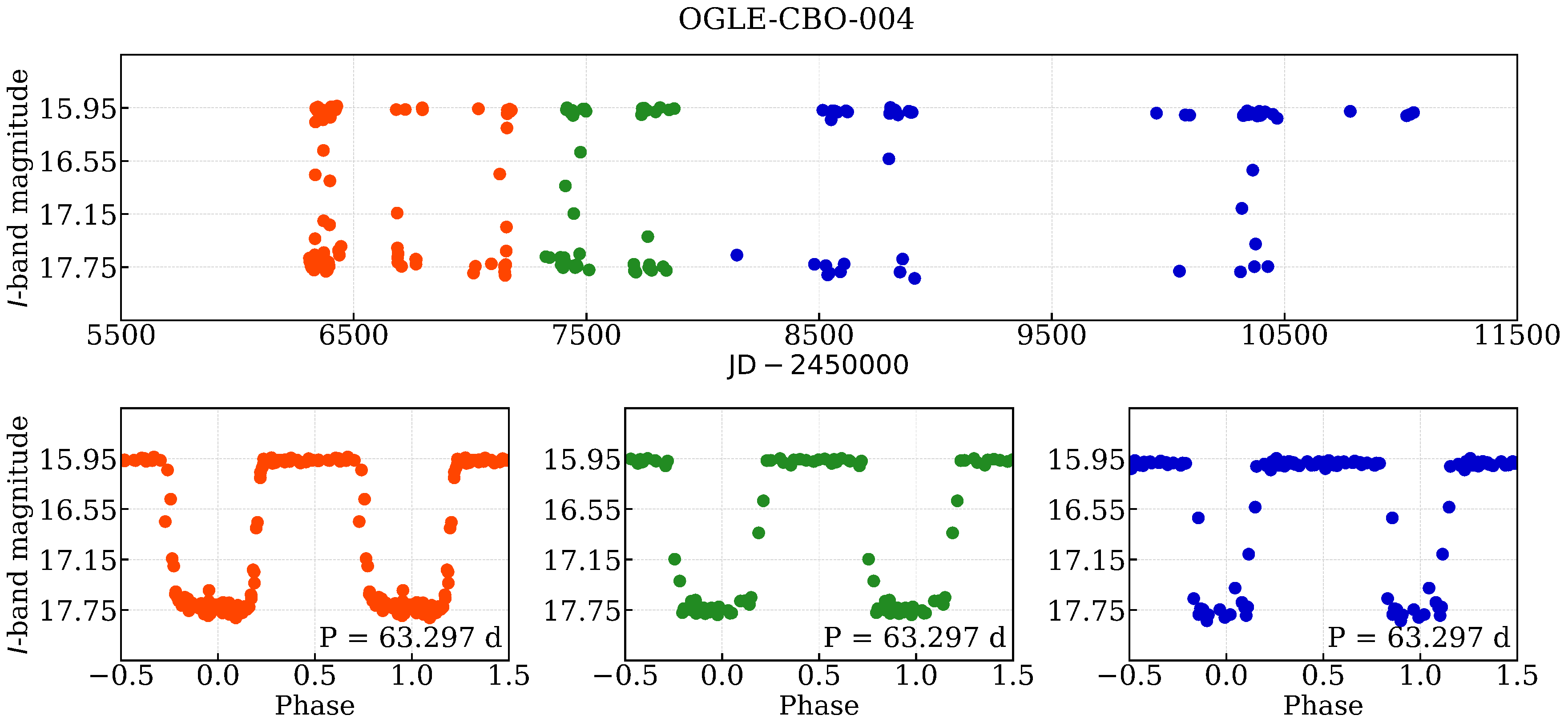}\\[1.cm]
\includegraphics[width=1.0\textwidth]{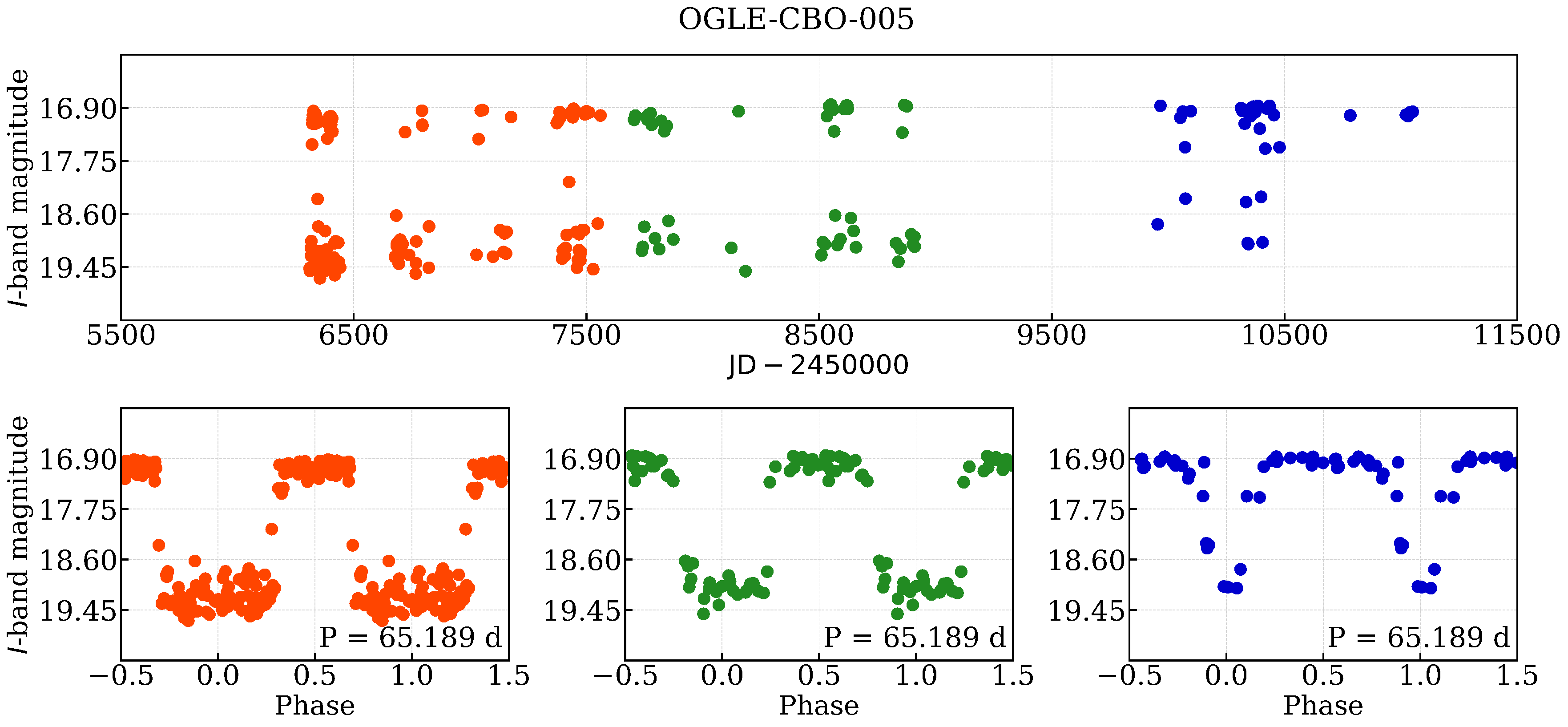}\\[1.cm]
\includegraphics[width=1.0\textwidth]{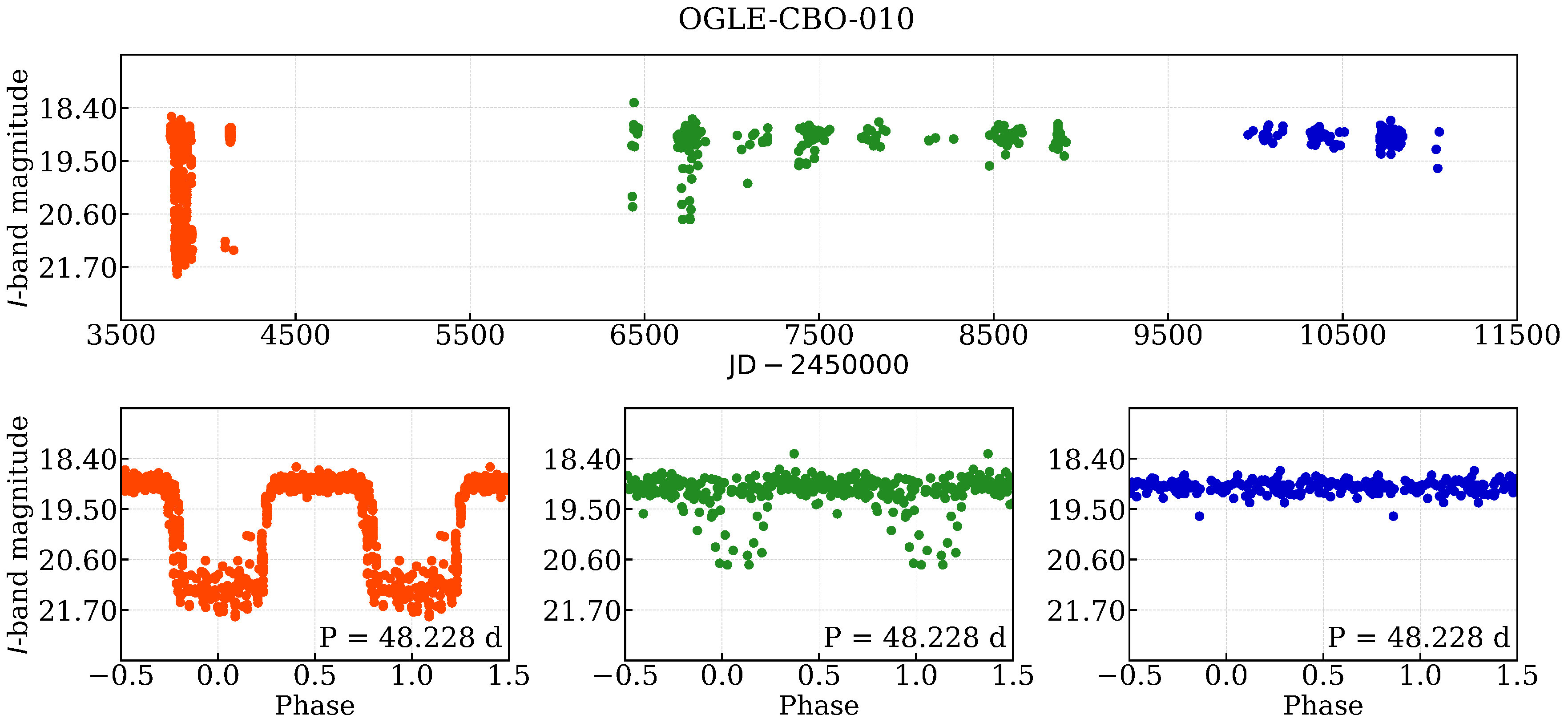}
\FigCap{$I$-band light curves of \textit{OGLE-CBO-004}, \textit{OGLE-CBO-005}, and \textit{OGLE-CBO-010}. For each star, three panels with phase-folded light curves are shown, where the colors correspond to the time intervals indicated on the time-series photometry light curve. The stars exhibit changes in the shape of their light curves over the years.}
\label{fig:fig9}
\end{figure}

\begin{figure}[p]
\includegraphics[width=1.0\textwidth]{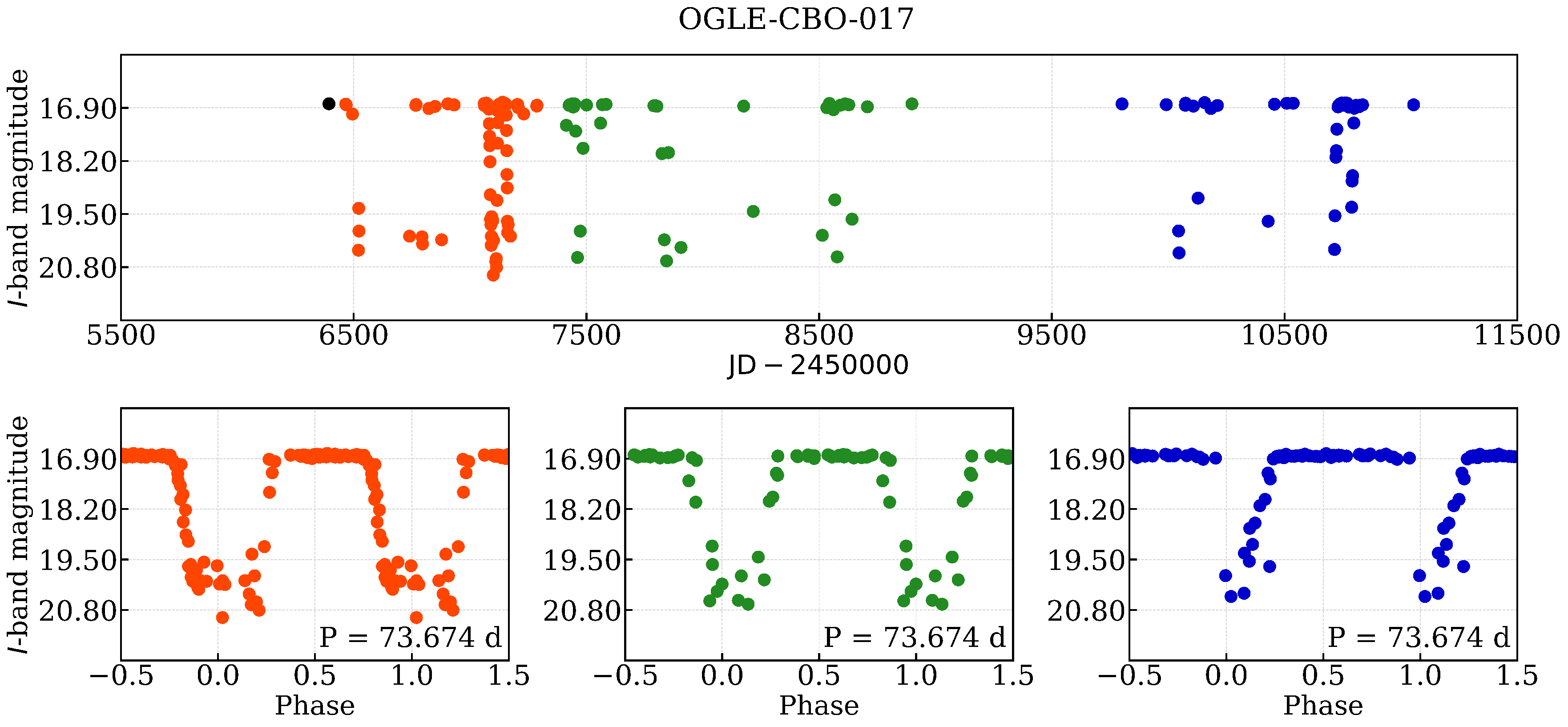}\\[1.cm]
\includegraphics[width=1.0\textwidth]{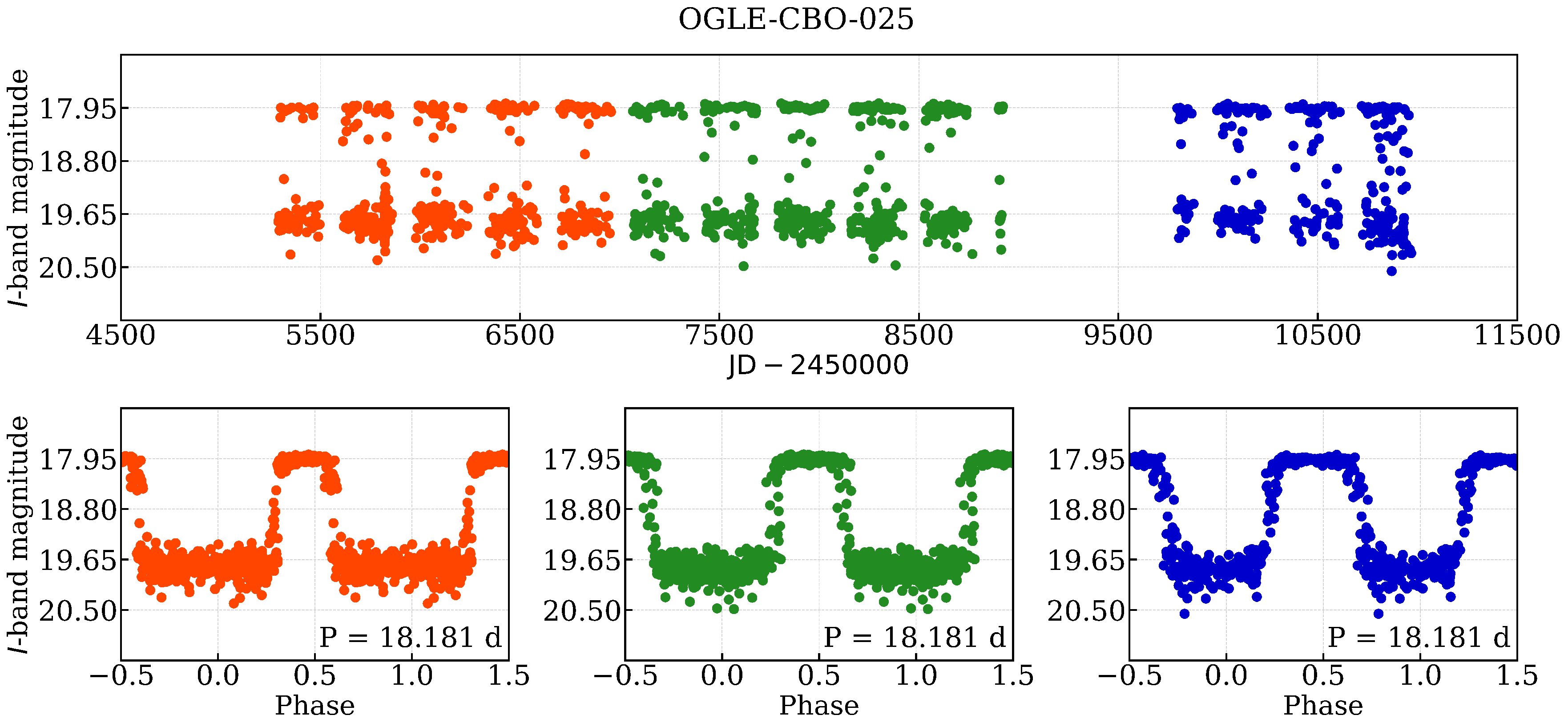}\\[1.cm]
\includegraphics[width=1.0\textwidth]{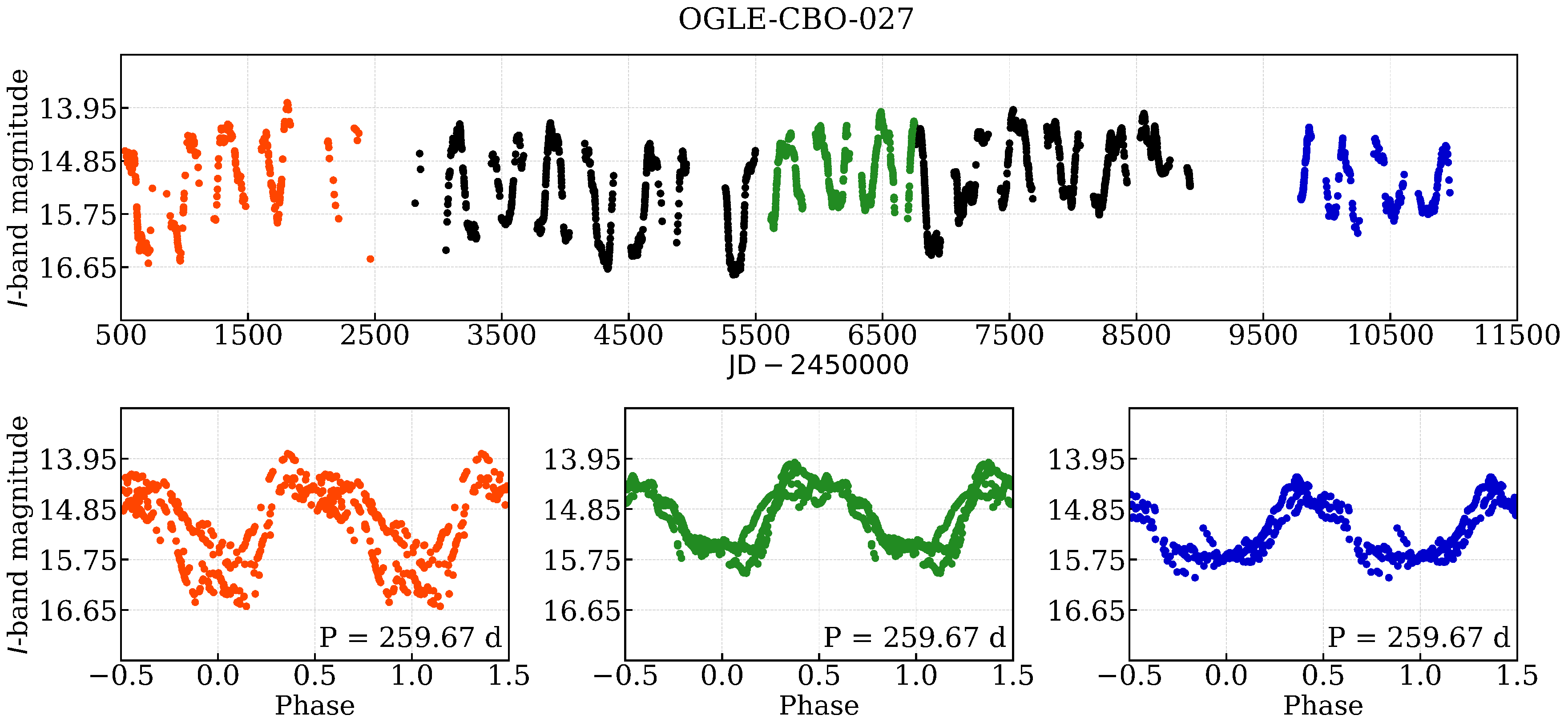}
\FigCap{$I$-band light curves of \textit{OGLE-CBO-017}, \textit{OGLE-CBO-025}, and \textit{OGLE-CBO-027}. For each star, three panels with phase-folded light curves are shown, where the colors correspond to the time intervals indicated on the time-series photometry light curve. The stars exhibit changes in the shape of their light curves over the years.}
\label{fig:fig10}
\end{figure}

\begin{figure}[t]
\includegraphics[width=1.0\textwidth]{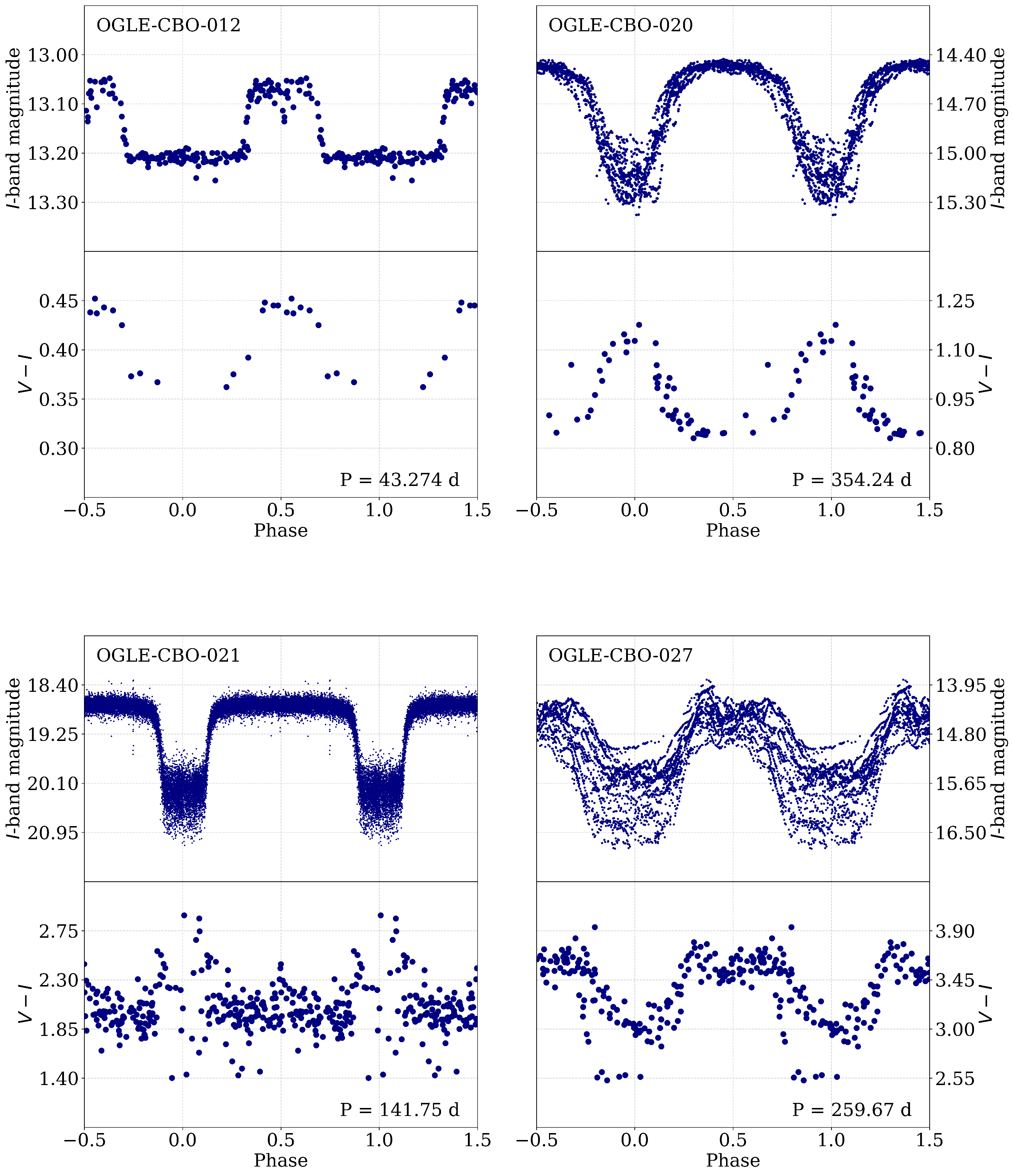}
\FigCap{Phase-folded {\it I}-band light curves (upper panels) and $V-I$ color curves (lower panels) for four representative CBO systems. Note the variation in stellar color with orbital phase.}
\label{fig:fig11}
\end{figure}

\subsection{Color Variations During the Occultations}
For stars with a sufficiently large number of OGLE observations in the $V$ band, we investigated the variation of the $V-I$ color as a function of orbital phase. Four examples are presented in Fig.~11. Majority of CBO systems become redder during eclipses (\eg OGLE-CBO-021), however some of them exhibit opposite behavior and become bluer (\eg OGLE-CBO-012 and OGLE-CBO-027). We suggest that this effect arises because, during occultation, the disk obscures the component with a significantly lower effective temperature, causing the hotter component to dominate the observed flux. We also show the light and color curves of OGLE-CBO-022, which exhibits a relatively large scatter in $V-I$ at minimum brightness. The large scatter is influenced by the very low observed brightness of the star, especially at minimum light, which results in large uncertainties in the $V-I$ color measurements. However, an additional cause may also be present. From the example of KH~15D, we know that just before complete occultation, one of the components illuminates the disk, which may preferentially scatter shorter-wavelength light. The light scattered by the disk contributes to the total observed flux, causing the stellar color at certain phases to shift toward the blue (Arulanantham \etal 2016; Aronow \etal 2018). In the case of OGLE-CBO-021, this effect would require further investigation; however, the remaining stars in our catalog for which we have a sufficient number of $V$-band observations do not show evidence of a similar phenomenon.

\subsection{OGLE-CBO-021 and OGLE-CBO-024 spectroscopy}
In Fig.~12, we show low-resolution spectra obtained for objects OGLE-CBO-021 and OGLE-CBO-024. In the case of OGLE-CBO-021, the spectrum was taken in the maximum light, while in the case of OGLE-CBO-024 it was close to its minimum light. Due to low brightness of the target stars in the optical range, we are able to spot only a few features in the spectra. Object OGLE-CBO-021 shows an evident H$\alpha$ line in emission, which is typical for young stellar objects. Unfortunately, we cannot conclude on the spectral type. Star OGLE-CBO-024 exhibits TiO bands which are characteristic for M type stars. A mild H$\alpha$ line in emission is also present.

\begin{figure}[h]
\includegraphics[width=1.0\textwidth, trim=0.0cm 11.5cm 0.0cm 4.5cm, clip]{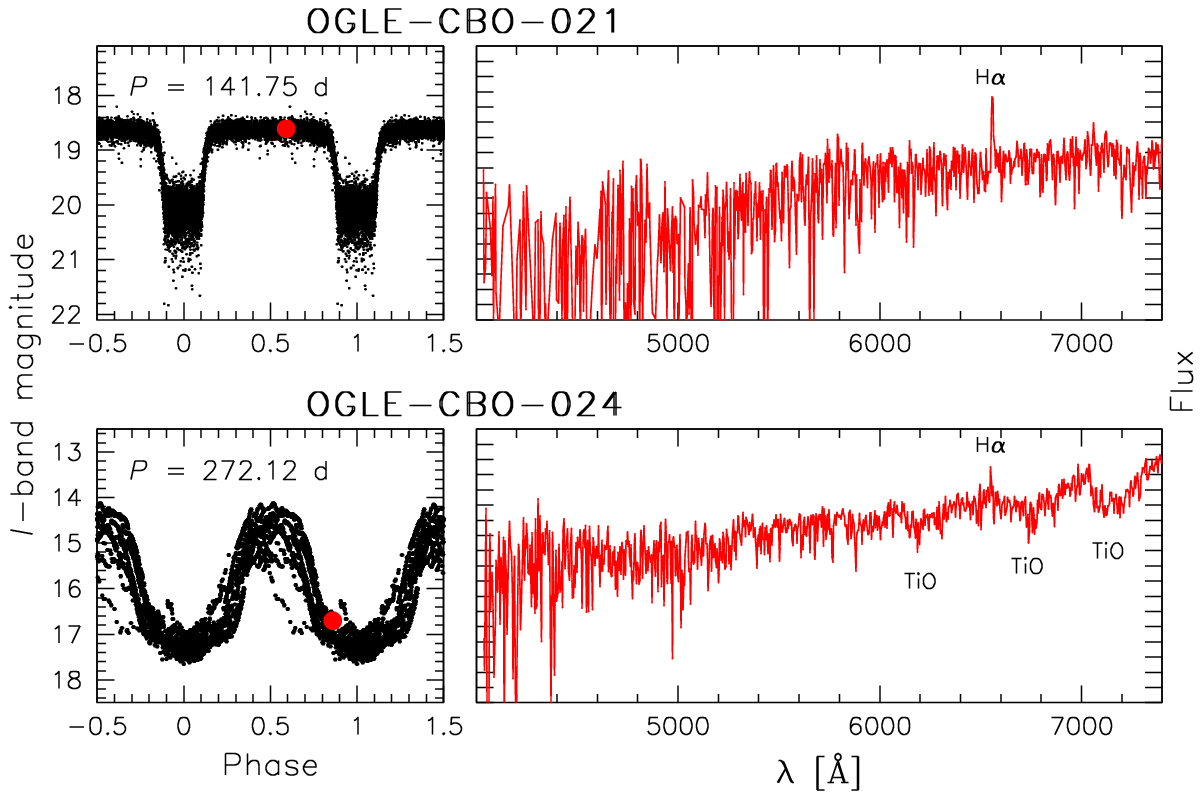}
\FigCap{Low-resolution spectra of objects OGLE-CBO-021 (top) and OGLE-CBO-024 (bottom)
together with the marked moment of the data acquision (as large red dots) in the
phase-folded light curves (on the left).}
\label{fig:fig12}
\end{figure}

\section{Summary}

In this work, we present 30 stars exhibiting rarely observed brightness variations. We suggest that their nature may be similar to that of KH~15D. Prior to this study, only 15 objects of this type were known (Hu \etal 2026), whereas here we report as many as 25 new candidates for this class.

Thanks to the systematic observations conducted within the OGLE project, we have obtained long-term light curves in the $I$ and $V$ bands, which are now becoming publicly available. These data enable us to trace a variety of changes in these objects, including variations in eclipse width, changes in amplitude, and other modifications of the light-curve morphology. We also briefly discuss selected systems and their properties.

\Acknow{This work has been funded by the National Science Centre, Poland, grant no.~2022/45/B/ST9/00243. For the purpose of Open Access, the author has applied a CC-BY public copyright license to any Author Accepted Manuscript (AAM) version arising from this submission. This work is based on observations collected at the European Southern Observatory under ESO programme 105.20EF.001.}


\begin{references}

\refitem{Agol, E., Barth, A.J., Wolf, S., and Charbonneau, D.}{2004}{\ApJ}{600}{781}

\refitem{Alard, C., and Lupton, R.H.}{1998}{\ApJ}{503}{325}

\refitem{Arulanantham, N.A. , Herbst, W., Cody, A.M., \etal}{2016}{\AJ}{151}{90}

\refitem{Aronow R.A., Herbst, W., Hughes, A. M., Wilner, D.J., and Winn, J.N.}{2018}{\AJ}{155}{47}

\refitem{Bernhard, K., Lloyd, C., and Huemmerich, S.}{2013}{Peremennye Zvezdy Prilozhenie}{13}{3}

\refitem{Bernhard, K., and Lloyd, C.}{2024}{\AA}{688}{58}

\refitem{Bernhard, K., Frank, P., Moschner, W., and Reffke, U.}{2024}{BAV Journal}{92}{1}

\refitem{Buzzoni, B., Delabre, B., Dekker, H., \etal}{1984}{The Messenger}{38}{9}

\refitem{ Castelli, F. and, Kurucz, R.L.}{2003}{IAUS}{210}{20}

\refitem{Chiang, E.I., and Murray-Clay, R.A.}{2004}{\ApJ}{607}{913}

\refitem{Evans, D.W., Irwin, M.J., and Helmer, L.}{2002}{\AA}{395}{347}

\refitem{Gaia Collaboration, Vallenari, A., Brown, A.G.A., \etal}{2023}{\AA}{674}{1}

\refitem{Groenewegen, M.A.T., and Blommaert, J.A.D.L.}{2005}{\AA}{443}{143}

\refitem{Hackstein, M., Fein, Ch., Haas, M., \etal}{2015}{\AN}{336}{590}

\refitem{Hamilton, C.M., and Herbst, W.}{2001}{\ApJ}{554}{201}

\refitem{Hamilton, C.M., Herbst, W., Vrba, F.J., \etal}{2005}{\AJ}{130}{1896}

\refitem{Hu, Z., Zhu, W. , Dai, F., \etal}{2024}{\ApJ}{977}{L28}

\refitem{Hu, Z., Zhu, W., Wang, S., and Wang, S.X.}{2026}{arXiv e-prints}{}{arXiv:2601.16828}

\refitem{Hunt, E.L., and Reffert, S.}{2023}{\AA}{673}{114}

\refitem{Kearns, K.E., and Herbst, W.}{1998}{\AJ}{116}{261}

\refitem{Kharchenko, N.V., Piskunov, A.E., Schilbach, E., R\"oser, S., and Scholz, R.D.}{2013}{\AA}{558}{53}

\refitem{Lucas, P. W. , Smith, L.C., Guo, Z., \etal}{2024}{\MNRAS}{528}{1789}

\refitem{Matsunaga, N., Fukushi, H., and Nakada, Y.}{\MNRAS}{2005}{364}{117}

\refitem{Mr\'oz, P., Pietrukowicz, P., Poleski, R., \etal}{2013}{\Acta}{63}{135}

\refitem{Mr\'oz, P., Udalski, A., Poleski, R., \etal}{2015}{\Acta}{65}{313}

\refitem{Pietrukowicz, P., Mr\'oz, P., Soszy\'nski, I., \etal}{2013}{\Acta}{63}{115}

\refitem{Plavchan, P., Gee, A.H., Stapelfeldt, K., and Becker, A.}{2008}{\ApJ}{684}{37}

\refitem{Plavchan, P. , G\"uth, T., Laohakunakorn, N., and Parks, J.R.}{2013}{\AA}{554}{110}

\refitem{Poon, M., Zanazzi, J.J., and Zhu, W.}{2021}{\MNRAS}{503}{1599}

\refitem{Rodr\'iguez-Ledesma, M.V., Mundt, R., Ibrahimov, M., \etal}{2012}{\AA}{544}{112}

\refitem{Rodr\'iguez-Ledesma, M.V., Mundt, R., Pintado, O., \etal}{2013}{\AA}{551}{44}

\refitem{Soszy\'nski, I., Udalski, A., Szyma\'nski, M.K., \etal}{2013}{\Acta}{63}{21}

\refitem{Soszy\'nski, I., Pawlak, M., Pietrukowicz, P., \etal}{2016}{\Acta}{66}{405}

\refitem{Soto, A.G., Ali, A., Newmark, A. \etal}{2020}{\AJ}{159}{135}

\refitem{Tody, D.}{1986}{Proc. SPIE}{627}{733}

\refitem{Tody, D.}{1993}{ASPC}{52}{173}

\refitem{Udalski, A., Kubiak, M., and Szyma\'nski, M.}{1997}{\Acta}{47}{319}

\refitem{Udalski, A., Szyma\'nski, M.K., Soszy\'nski, I., and Poleski, R.}{2008}{\Acta}{58}{69}

\refitem{Udalski, A., Szyma\'nski, M.K., and Szyma\'nski, G.}{2015}{\Acta}{65}{1}

\refitem{Udalski, A., Soszy\'nski, I., Pietrukowicz, P., \etal}{2018}{\Acta}{68}{315}

\refitem{Watson, C.L., Henden, A. A., and Price, A.}{2006}{SASS}{25}{47}

\refitem{Windemuth, D., and Herbst, W.}{2014}{\AJ}{147}{9}

\refitem{Winn, J.N., Holman, M.J., Johnson, J.A., Stanek, K.Z., and Garnavich, P.M.}{2004}{\ApJ}{603}{45}

\refitem{Wrona, M., Ratajczak, M., Ko\l{}aczek-Szyma\'nski, P.A., \etal}{2022}{\ApJS}{259}{16}

\refitem{Zhu, W., Bernhard, K., Fang, M., \etal}{2022}{\ApJ}{933}{21}

\end{references}
\end{document}